\documentclass[referee]{raa}  
\usepackage{natbib}
\usepackage{graphicx,times}             %for PS/EPS graphics inclusion, new
\usepackage{amssymb,amsmath,mathrsfs}
\bibpunct{(}{)}{;}{a}{}{,}
\usepackage[pagebackref=true]{hyperref}

\newcommand{\bea}{\begin{eqnarray}}
\newcommand{\eea}{\end{eqnarray}}

\newcommand{\Tt}{\mathcal{T}}
\newcommand{\Gg}{\mathcal{G}}
\newcommand{\Pp}{\mathcal{P}}
\newcommand{\Ff}{\mathcal{F}}

\newcommand{\Oo}{\mathcal{O}}
\newcommand{\Ww}{\mathcal{W}}
\newcommand{\bB}{\mathbb{B}}
\newcommand{\bN}{\mathbb{N}}
\newcommand{\sP}{\mathscr{P}}

\begin{document}

\title{How to coadd images: II. Anti-aliasing and PSF deconvolution}

 \volnopage{ {\bf 2023} Vol., {\bf X} No. {\bf XX}, 000--000}
 \setcounter{page}{1}
 
\author{Lei Wang\inst{1}\inst{4}\inst{5}\thanks{E-mail: leiwang@pmo.ac.cn}, Huanyuan Shan\inst{3}\inst{10}\inst{11}, Lin Nie\inst{2}\inst{4}, Dezi Liu\inst{6}\inst{7}\inst{11}, Zhaojun Yan\inst{3}, Guoliang Li\inst{1}\inst{4}\inst{5}, Cheng Cheng\inst{8}\inst{9}, Yushan Xie\inst{3}, Han Qu\inst{1}\inst{4}\inst{5}, Wenwen Zheng\inst{1}\inst{4}\inst{5}, Xi Kang\inst{5}}
  \institute{
   $^1$ Purple Mountain Observatory, Chinese Academy of Sciences, No. 10 Yuan Hua Road, Nanjing 210023, China;\\
   $^2$ Department of Information Engineering, Wuhan Institute of City, Wuhan, Hubei 430083, China;\\
   $^3$ Shanghai Astronomical Observatory, Nandan Road 80, Shanghai 200030, China;\\
   $^4$ National Basic Science Data Center, Building No.2, 4, Zhongguancun South 4th Street, Haidian District, Beijing 190, China;\\
   $^5$ Zhejiang University-Purple Mountain Observatory Joint Research Center for Astronomy, Zhejiang University, Hangzhou 327, China;\\
   $^6$ South-Western Institute for Astronomy Research, Yunnan University, Kunming, 650500, China;\\
   $^7$ The Shanghai Key Lab for Astrophysics, Shanghai Normal University, Shanghai, 200234, China;\\
   $^8$ Chinese Academy of Sciences South America Center for Astronomy, National Astronomical Observatories, CAS, Beijing 100101, China;\\
   $^9$ CAS Key Laboratory of Optical Astronomy, National Astronomical Observatories, Chinese Academy of Sciences, Beijing 100101, China;\\
   $^{10}$ Key Laboratory of Radio Astronomy and Technology, Chinese Academy of Sciences, A20 Datun Road, Chaoyang District, Beijing, 100101, China;\\
 $^{11}$ University of Chinese Academy of Sciences, Beijing 100049, China
  } 
   
\vs \no
 {\small Received 2023 October XX; accepted XXXX July XX}

%\date{{\sc Draft: } \today }
%zhaojunyan@shao.ac.cn
%\pagerange{\pageref{firstpage}--\pageref{lastpage}} \pubyear{2013}
%\maketitle
%\label{firstpage}

%%%%%%%%%%%%%%%%%%%%%%%%%%%%%%%%%%%%%%%%%%%%%%%%%%%%%%%%%%%%%%%%%%%%%%%%%%

\abstract{
We have developed a novel method for co-adding multiple under-sampled images that combines the iteratively reweighted least squares and divide-and-conquer algorithms. Our approach not only allows for the anti-aliasing of the images but also enables PSF deconvolution, resulting in enhanced restoration of extended sources, the highest PSNR, and reduced ringing artefacts. To test our method, we conducted numerical simulations that replicated observation runs of the CSST/VST telescope and compared our results to those obtained using previous algorithms. The simulation showed that our method outperforms previous approaches in several ways, such as restoring the profile of extended sources and minimizing ringing artefacts. Additionally, because our method relies on the inherent advantages of least squares fitting, it is more versatile and does not depend on the local uniformity hypothesis for the PSF. However, the new method consumes much more computation than the other approaches. 
}

 \authorrunning{Wang Lei et al. }    %author_head in even pages
 \titlerunning{How to coadd images}  % title_head in odd pages
 \maketitle
%-------------------------------------------------------------------------%
\section{Introduction}
\label{sect:intro}
%-------------------------------------------------------------------------%

As technology has advanced, computer software has become increasingly important in the field of image processing. To simplify data processing, an Image Acquisition System (IAS) is often utilized for image digitization. Essentially, an IAS acts as a digitizer that converts continuous signals into digital ones by recording data with detectors, such as pixels, which appear in a mosaic-like pattern known as pixelation. However, due to the diffraction limit of the optics equipment, the images captured by the IAS are band-limited. This means that the signal recorded by the image has a maximum spatial frequency or resolution, as described in \cite{Fruchter+2011}. A band-limited signal can be fully reconstructed by high-sampling imaging, according to the Nyquist sampling theorem. Nevertheless, many astronomical images do not meet the Nyquist sampling criterion due to technical or economic limitations, resulting in under-sampled images(\citealt{Fruchter+2002}). 

Under-sampled images suffer from an aliasing effect, which blurs all signals below a sampling interval. To achieve a resolving power that approaches the diffraction limit, the sampling rate needs to be increased. This can be accomplished by combining multiple exposures using various coaddition methods, such as {\it shift-and-add} (\citealt{Bates+1980, Farsiu+2004a}), {\it Drizzle} (\citealt{Fruchter+2002}), {\it Super-Drizzle} (\citealt{Takeda+2006}), {\it IMCOM} (\citealt{Rowe+2011}), {\it iDrizzle} (\citealt{Fruchter+2011}), {\it SPRITE} (\citealt{Mboula+2015}), and {\it fiDrizzle} (\citealt{Wang+2017}), as well as iterative back-projection ({\it IBP}, \citealt{Irani+1993,Symons+2021}). {\it Drizzle} has become the standard for combining images taken by the Hubble Space Telescope (HST) and the James Webb Space Telescope (JWST). Some Drizzle-based methods are widely used to restore fine details of under-sampled multi-exposures and fuse images from different equipment. {\it SPRITE} and {\it IBP} are developed to reconstruct the Point Spread Function (PSF) using stars in even one exposure.

Due to the diffraction of light, the image of a point-like source in the image plane is not actually a point. Instead, it appears as an extended object, known as the PSF effect. This degradation of the observed image is caused by the convolved PSF, but PSF deconvolution technology can improve resolution by compensating for this numerically. Along with resolution improvement, PSF deconvolution also enhances contrast and reduces noise(\citealt{Sage+2017}). There are various algorithms for PSF deconvolution(\citealt{Starck+2002}), including the Fourier-quotient method, {\it CLEAN} method(\citealt{Hogbom+1974}), Bayesian approach[including {\it Landweber} method(\citealt{Landweber+1951}), {\it Richardson-Lucy} algorithm(\citealt{Richardson+1972,Lucy+1974,Shepp+1982}) ], wavelet-based deconvolution, and super-resolution techniques(\citealt{Gerchberg+1974, Hunt+1994, Lauer+1999,Elad+1999,Capel+2003,Park+2003,Farsiu+2004b,Ouwerkerk+2006,Tian+2011,Nasrollahi+2014,Yue+2016,Symons+2021}). The PSF deconvolution has numerous benefits, such as its applicability to even the simplest optical setup, reduction of financial costs, and streamlining of the acquisition pipeline. However, when noise contaminates the PSF-convolved image, PSF deconvolution becomes an ill-posed problem(\citealt{Starck+2002,Sage+2017}). Regularized methods are often employed to generate an approximation(\citealt{Takeda+2007,Ng+2007,Takeda+2009,Yuan+2010,Babacan+2011,Su+2012,Liu+2014,Zhang+2012}), which is effective and flexible in reducing noise amplification, ringing effect, and flux divergence.

However, most previous works aim to either the PSF deconvolution for single exposure (e.g. {\it CLEAN}, {\it Landweber}, {\it Richardson-Lucy}, maximum entropy algorithm, etc.) or to multiple exposures coaddition but without PSF deconvolution (e.g.  {\it shift-and-add} and {\it Drizzle} based methods). Only a few works mention the under-sampled multi-exposures coaddition (MEC) with PSF deconvolution, e.g. {\it Stark-Pantin}: Eq. 53 in \cite{Starck+2002} and {\it UPDC}: Eq. 10 in \cite{Wang+2022}. In this paper, we systematically study how to achieve superresolution in the iterative MEC by anti-aliasing and PSF deconvolution coaddition (AAPD). We test several recovery methods with numerical simulations.

  %-------------------------------------------------------------------------%
  \section{Imaging model and the least squares}\label{sec_LSA}
  %-------------------------------------------------------------------------%
The observation image records not only the shape or flux from the objects of study but also a set of combined observational effects, e.g. blurring effects (PSF, seeing, vignetting, etc.), sampling effect (pixelation, image field distortion), noise effect (equipment noise, environment noise), etc. Therefore, a reasonable imaging model should take those observational effects into account.

Let $\bB$ be an image degrading operator (i.e. blurring) which represents a combination of a series of image operations e.g. PSF convolution, pixelation, etc. The formation of an under-sampled observation image $\Gg$ with the size of $U \times V$ pixels can be formally expressed as
\begin{equation}\label{model}
\Gg = \bN \left\{ \bB\{\Oo \} \right\},
\end{equation}
where $\Oo$ is the intrinsic, continuous surface brightness of the sky. Assuming that $\Oo$ contains a fine grid of size $M\times N$, we can represent the downgraded effects of the image, except for the noise, using $\bB\{\Oo\}$. The noise contamination on the downgraded image is described by $\bN$. To recover the original image $\Oo$ from $\Gg$, we need to know the combined effects of $\bN\{{\bB} \{\cdot\}\}$ in Eq.(\ref{model}) in advance. In this study, we focus on the recovery method and assume that the PSFs and the position shift on all dithered exposures are well measured beforehand. Table~\ref{params} provides a list of important symbols and their representations.

\begin{figure}
\centering
\includegraphics[width=4.1in]{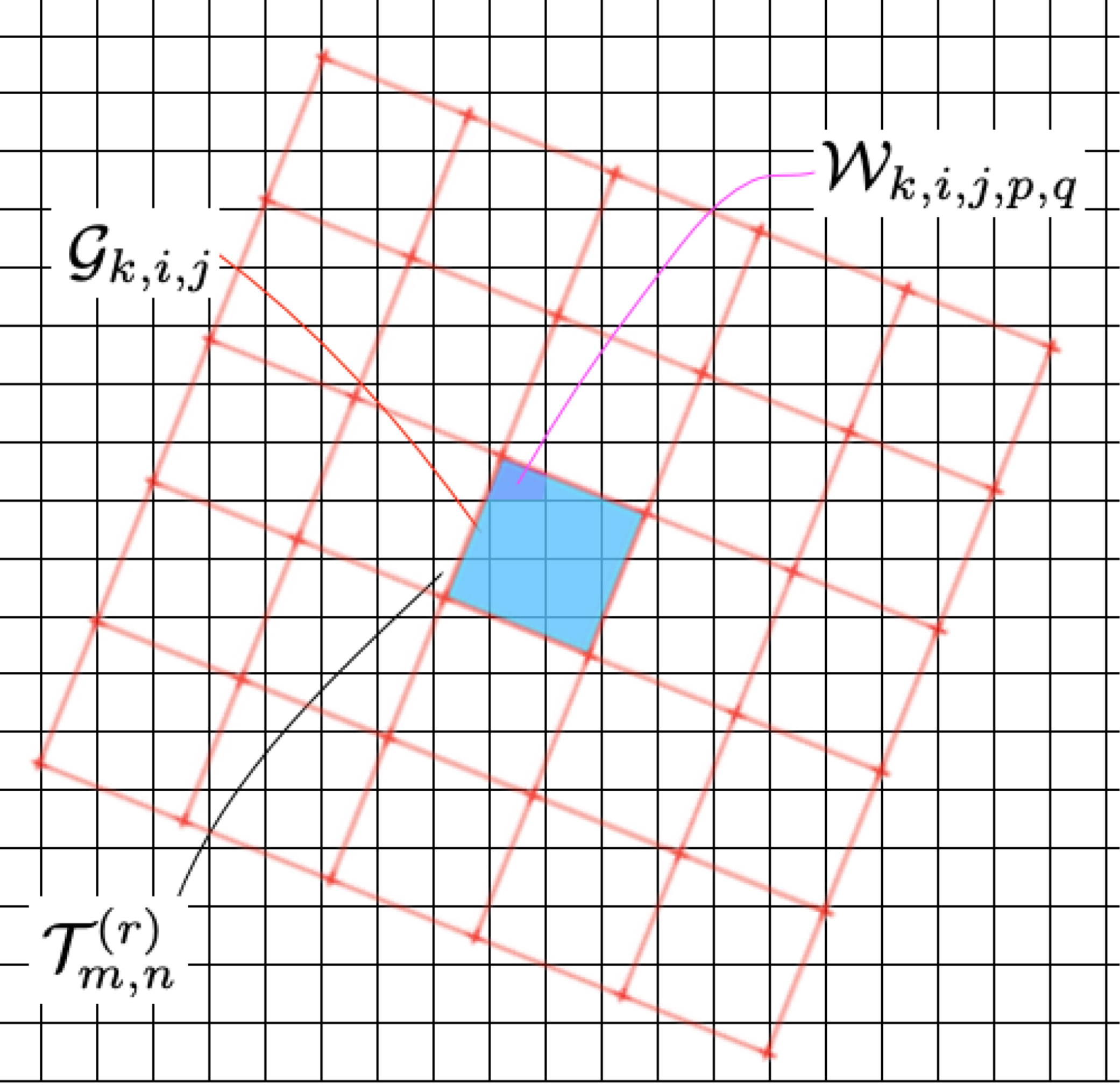} 
\caption{Resampling mechanism. The target grid $\Tt^{(r)}_{m,n}$ is shown as the fine (in black), the observation pixels $\Gg_{k,i,j}$ are shown as coarse (in red). The flux weight $\Ww_{k,i,j,p,q}$ (in pink) is determined by the clipped area between the observation pixels and the target grid. Here the $TS$ space ($p,q$) includes all target cells that overlap with the observation pixel $\Gg_{k,i,j}$ (blue region).}\label{figz}
\end{figure}

Supposing that we have $L$ blurred (PSF convolved and/or under-sampled) exposures $\{\Gg_1, \Gg_2... \Gg_L\}$\footnote{Frame $\Gg_k$ ($k=1 ... L$) is obtained from the $k$-th observation} with $U \times V$ pixels for the same object and the known blurring effects $\bB \{\cdot\}$, can we mock another set of frames $\{\Ff_1, \Ff_2... \Ff_L\}$ to approach the observations, then constrain the original image? To answer this question, we try to use the least squares to calculate a $\chi^2$ for all pixels between the observation and mock samples. Theoretically, the $\chi^2$ can be represented as
\bea 
\chi^2 =\sum^{L,U,V}_{k,i,j}\frac{(\Ff_{k,i,j}-\Gg_{k,i,j})^2}{\theta_{k,i,j}^2},
\label{eq1}
\eea
where $\Ff={\bB} \{\Tt\}$ and $\Tt$ is a target image with size of $M\times N$ fine grids, i.e. the estimation of the original image $\Oo$. While $\theta_{k,i,j}$ is the noise at the pixel of the $i$-th row and $j$-th column of the $k$-th observational frame.  Note that the summation of the squares is over $L\times U\times V$ pixels. Since the blurring effects $\bB \{\cdot\}$ are known, minimizing the $\chi^2$ results in an estimation of the target image $\Tt$.
By differentiating $\chi^2$ with respect to $\Tt$ and setting the derivative to zero, we have
\bea 
\frac{\partial \chi^2}{\partial \Tt}=0 .
\label{eq2}
\eea
It is very difficult to solve the equation Eq.~\ref{eq2} directly, especially for huge amounts of pixels. Actually, Eq.~\ref{eq2} is a system of linear equations with $M\times N$ unknowns and $L\times U\times V$ conditions. As a kind of fallback solution, we employ a divide-and-conquer algorithm to solve Eq.~\ref{eq2} on each target grid ($m,n$) at a time. The $Drizzled$ grids are used as the initial values. The target grid ($m,n$) will be renewed after solving Eq.~\ref{eq2}. The program checks if the output meets the preset threshold\footnote{In this work, we suggest using the $\chi^2$ as the threshold.} when all grids are updated. If not, the updated grids will take the place of the initial values in Eq.~\ref{eq2}. Then it repeats the above steps until the updated grids meet the threshold. Therefore, Eq.~\ref{eq2} is solved in an iterative approach like the following:
\bea 
\sum^{OS}_{k,i,j} \frac{1}{\theta_{k,i,j}^2}\sum^{TS}_{p,q}\Ww_{k,i,j,p,q}\big{(}\sum^{TS}_{s,t}\Pp_{k,p,q,s,t}\Tt^{(r)}_{s,t} +(\Tt^{(r+1)}_{m,n}- \Tt^{(r)}_{m,n})\Pp_{k,p,q,m,n}\big{)}\sum^{TS}_{p,q}\Ww_{k,i,j,p,q}\Pp_{k,p,q,m,n}\\\nonumber
=\sum^{OS}_{k,i,j} \frac{\Gg_{k,i,j}}{\theta_{k,i,j}^2}\sum^{TS}_{p,q}\Ww_{k,i,j,p,q}\Pp_{k,p,q,m,n}
\label{eq3}
\eea
where $\Pp_k$ is a fine PSF measured from the $k$-th frame. The PSF must have the same sampling rate as the target image. $\Ww_{k,i,j,p,q}$ is a flux allocation weight that stands for how many areas of an observation pixel ($k,i,j$) are overlapped by the target pixel ($p,q$), shown as Figure \ref{figz}. In this work, the overlapped polygon of two grids is clipped using a polygon clipping algorithm {\it Clipper2}\footnote{https://github.com/AngusJohnson/Clipper2}. The area of the clipped polygon is calculated using the Green formula. Indexes $r$ and $r+1$ represent the sequence number of the iteration step. $OS$ is a space that includes all observation pixels that are overlapped with PSF $\Pp_k$ which is centred at the target pixel ($m,n$), while $TS$ includes all target pixels that are overlapped with the $OS$ observation pixels. 

Let $\Psi_{k,i,j,m,n}=\sum^{TS}_{p,q}\Ww_{k,i,j,p,q}\Pp_{k,p,q,m,n}$ and $\Phi_{k,i,j,m,n}^{(r)}=\sum^{TS}_{p,q}\Ww_{k,i,j,p,q}\sum^{TS}_{s,t}\Pp_{k,p,q,s,t}\Tt^{(r)}_{s,t}$, the Eq.~\ref{eq3} can be rewritten as
\bea 
\Tt^{(r+1)}_{m,n}=\Tt^{(r)}_{m,n}+\frac{\sum^{OS}_{k,i,j} \frac{\Psi_{k,i,j,m,n}}{\theta_{k,i,j}^2}(\Gg_{k,i,j}-\Phi^{(r)}_{k,i,j,m,n})}{\sum^{OS}_{k,i,j} \frac{\Psi_{k,i,j,m,n}^2}{\theta_{k,i,j}^2}}
\label{eq4}
\eea
which is an iterative solution to the least squares. $\Phi_{k,i,j,m,n}^{(r)}$ is the mimic flux of pixels which are overlapped by the target grid $\Tt^{(r)}_{m,n}$. In order to eliminate contaminations from cosmic rays or abnormal pixels, the weight $\theta_{k,i,j}$ is dynamically adjusted with the iterations, which results in robust regression of the algorithm. This is actually an application of the {\it Iteratively Reweighted Least Squares} (IRLS) method in image stacking. We call this divide-and-conquer IRLS method i.e. Eq.~\ref{eq4} as {\it Dirles}. The initial value of the iteration $\Tt^{(0)}$ is set as the $Drizzled$ real exposures to automatically introduce telescope effects such as vignetting, instrument noise, and saturation overflow. Therefore, these effects do not need to be simulated anymore in the {\it Dirles}. 

In literature, there are two algorithms besides the {\it Dirles} which can achieve the {\it anti-aliasing and PSF deconvolution coaddition} simultaneously. These algorithms are called {\it Starck-Pantin} (\citealt{Starck+2002}) and {\it UPDC} (\citealt{Wang+2022}). Both {\it Starck-Pantin} and {\it UPDC} are forward modelling methods that infer the original image by maximizing the likelihood between observations and mocks. The algorithms like {\it Richardson-Lucy}, {\it Starck-Pantin}, and {\it UPDC} use FFT to convolve the PSF. However, the {\it Dirles} algorithm convolves the PSF in the real space, making it more suitable for situations where the PSF changes dramatically. Although it consumes much more computation than the previous algorithms, it is a better choice for such situations.

Drizzling the multi-exposures of an ideal point source can form a blurring effect in the target image (grid), which originates from the pixelation. It is called pixelation blur to distinguish it from the ordinary PSF(see \citealt{Wang+2022}  for more details). The pixelation blur is not homogeneous, even not continuously changing in the whole target image. Because different point sources have different positions therefore different cases of pixels coaddition. Due to the heterogeneity of pixelation blur, theoretically, it is not a good choice to deconvolve the PSF by using ordinary methods e.g. {\it Landweber} or {\it Richardson-Lucy} algorithm on the $Drizzled$ image. 

\begin{figure}
\centering
\includegraphics[width=6.1in]{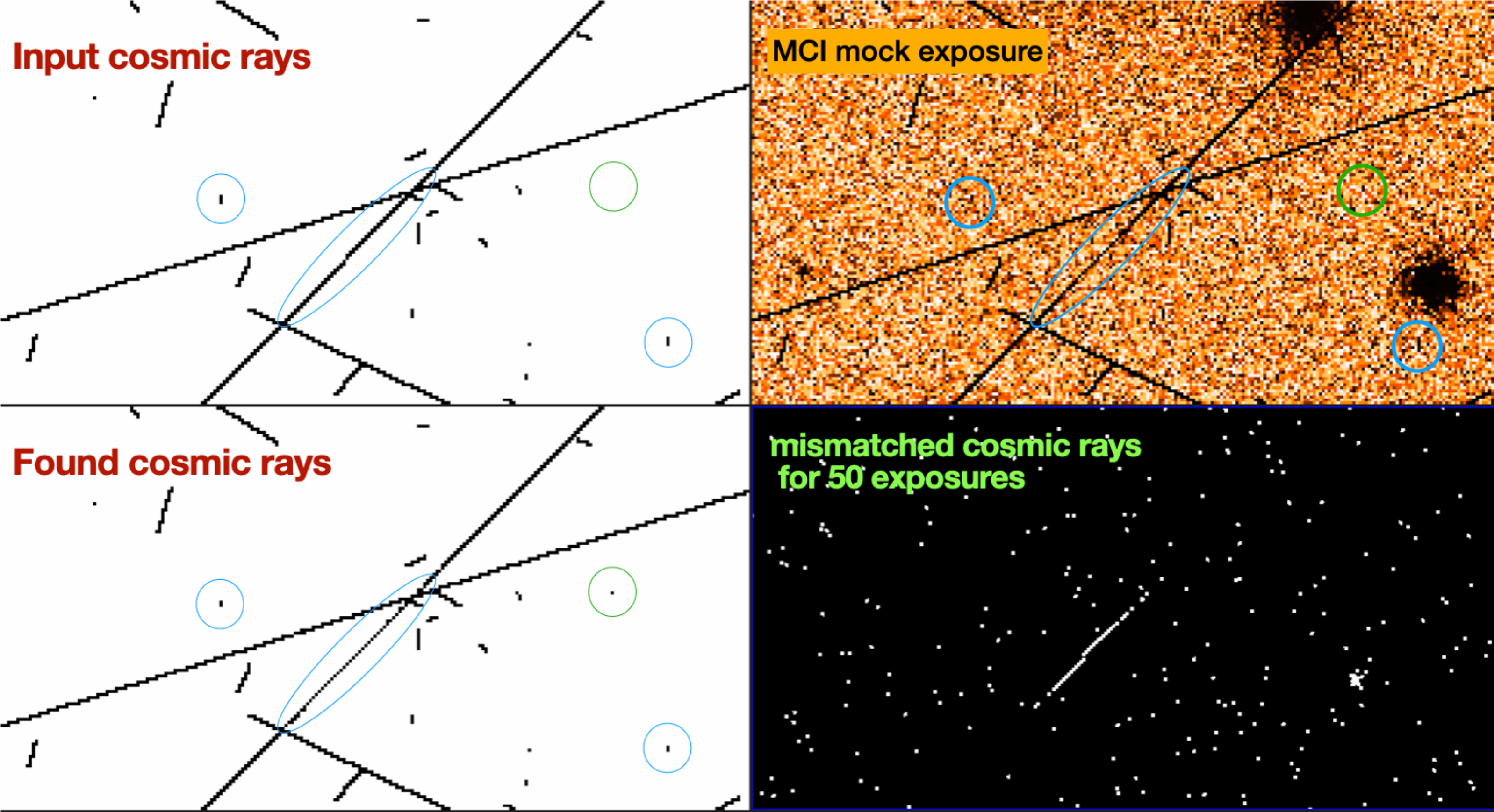} 
\caption{Cosmic-ray removal. For one of the exposures, there are 4 circles in the top panels and bottom left. The bottom right panel shows all the mismatched cases (including 50 exposures). Circles in green stand for the misidentification cases (They are not real cosmic rays), while blue for the omissions.}\label{cosmic-ray}
\end{figure}

\section{Image completion: cosmic rays and bad pixels replacement}
Instead of using the $Drizzle$ method for cosmic-ray removal(\citealt{Fruchter+2002}), we apply an improved statistical algorithm that identifies and replaces abnormal regions with reasonable values. The new algorithm can significantly reduce the mismatched cosmic rays ($\sim50\%$) in our {\it CSST-MCI} mock test, e.g. Figure \ref{cosmic-ray}. The process of image completion involves the following steps.

\begin{enumerate}
\item{To start, let's create a target grid, denoted by $\Tt_{m,n}$, with a 1:1 sampling rate. This grid will be utilized to resample all $L$ exposures $\Gg_{k,i,j}$ ($k=1 ... L$) based on the $WCS$ parameters specified in their headers.}

\item{For each target pixel located at position $(m, n)$, we collect a statistical sample of overlapping exposure pixels, denoted by $\mathbb{G}_0(m, n)$. We gather a total of $M\times N$ samples. These statistical samples are then utilized to calculate the median value, providing the first estimation of an image that is free of cosmic rays or bad pixels, known as $\Tt_{\rm mid}(m, n)$.}

\item{The mean, $\mu_0(m, n)$, and standard deviation, $\sigma_0(m, n)$, are also calculated for each sample $\mathbb{G}_0(m, n)$. We use these values to identify cosmic-ray candidates by checking if the exposure pixels with a flux of $\Gg_{k,i,j}> \Tt_{\rm mid}(m, n)+\nu_0\times\sigma_0(m, n)$ meet the preliminary voting condition. Here, $\nu_0$ is an artificial threshold set to 5 in our work. If a region occasionally has multiple cosmic rays, the mean $\mu_0(m, n)$ can be overestimated. Therefore, we replace the mean $\mu_0(m, n)$ with the median $\Tt_{\rm mid}(m, n)$ in the preliminary voting condition.}

\item{Typically, an exposure pixel, $\Gg_{k,i,j}$, is covered by 1 to 5 neighbouring target pixels that act as "voters" in determining whether a cosmic-ray candidate is present. If a cosmic-ray candidate receives all the votes of the voters in the district, it is flagged as a cosmic ray. We remove the cosmic rays from the sample $\mathbb{G}_0(m, n)$ to obtain a relatively pure sample, denoted by $\mathbb{G}_1(m, n)$.}

\item{Once again, we calculate the mean, $\mu_1(m, n)$, and standard deviation, $\sigma_1(m, n)$, for each sample $\mathbb{G}_1(m, n)$. We then perform a series of similar operations as described above, but with a refreshed condition $\Gg_{k,i,j}> \mu_1(m, n)+\nu_1\times\sigma_1(m, n)$, where $\nu_1$ is set to 5. This allows us to remove cosmic rays from sample $\mathbb{G}_1(m, n)$, resulting in a reduced sample called $\mathbb{G}_2(m, n)$. Please note that some exposure pixels with flux $\Gg_{k,i,j}> \mu_1(m, n)+\nu_0\times\sigma_1(m, n)$ are removed for their cosmic-ray identification, but those with flux $\Gg_{k,i,j}< \mu_1(m, n)-\nu_0\times\sigma_1(m, n)$ are not. This means that using the mean $\mu_2(m, n)$ and standard deviation $\sigma_2(m, n)$ taken from sample $\mathbb{G}_2(m, n)$ to generate a Gaussian distribution is biased. To avoid that issue, we fit the sample $\mathbb{G}_2(m, n)$ to a Gaussian profile using the non-linear least squares fitting code $\it MPFIT$\footnote{https://pages.physics.wisc.edu/~craigm/idl/cmpfit.html}.}
\item{Each bad pixel or cosmic ray exposure pixel will be replaced with a random value generated by its corresponding Gaussian profile. }

\end{enumerate}
The above method is not valid for pixels that are overlapped by only one or two exposures due to the lack of a statistical sample. These cases usually occur at the marginal area of the target image $\Tt_{m,n}$, and can be ignored. Figure \ref{cosmic-ray} shows a case for cosmic-ray removal. The circles and ellipticals denote the mismatched cosmic rays. The bottom right panel shows the total number of mismatched cases for 50 exposures. The blue circles and ellipticals represent cosmic rays that were omitted due to their low flux blending with the background, while the green circles indicate cases of mistaken removal caused by the threshold set-up, i.e. a small probability event below the threshold.

\begin{figure}
\centering
\includegraphics[width=6.1in]{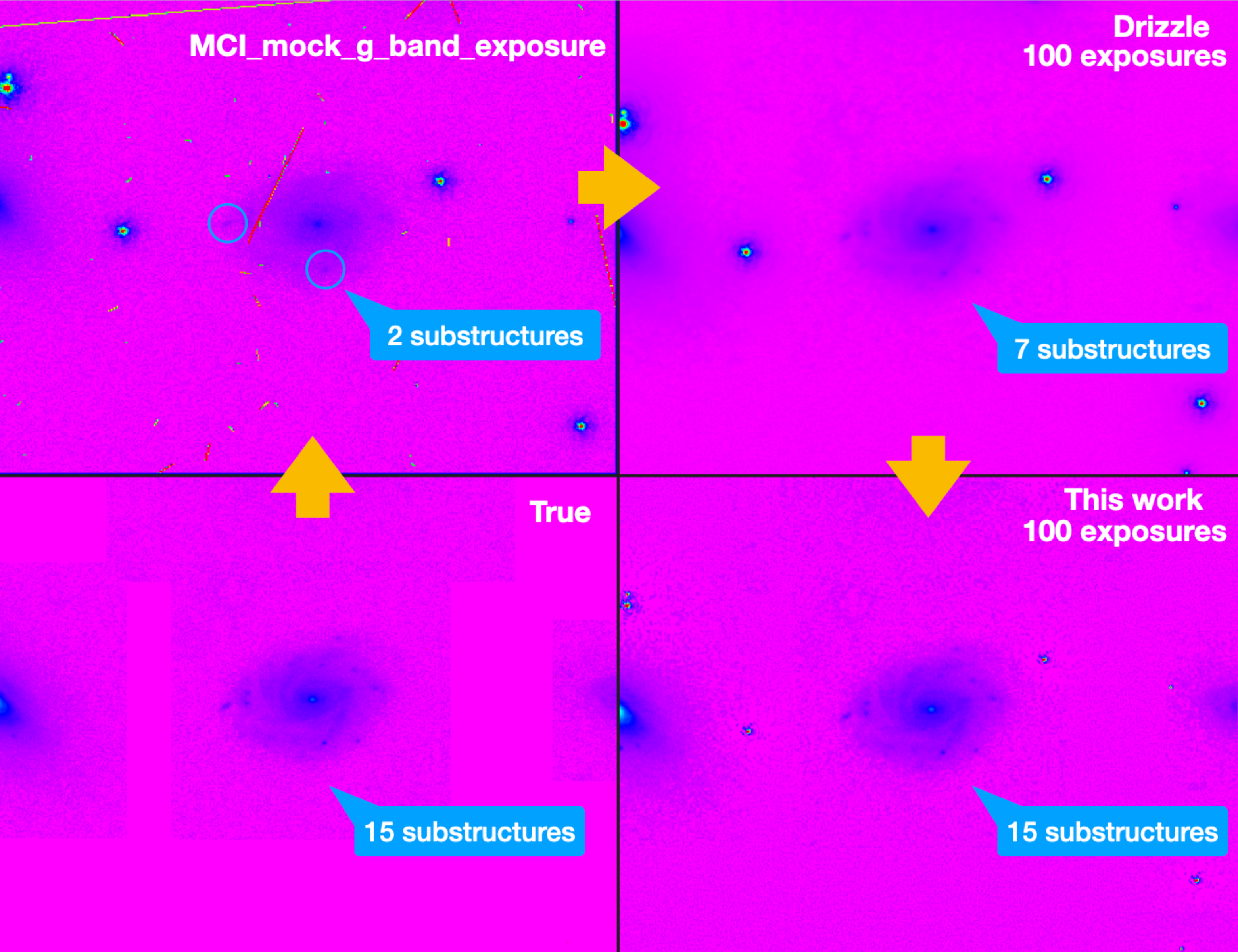} 
\caption{{\it CSST-MCI} mock and multi-exposure coaddtion. The bottom left panel shows the ground-truth image as the mock input, which is extracted from the HST observation with at least 15 substructures on the central galaxy. One of the $g$-band exposures with 2 obvious substructures is illustrated on the top left. There are about 7 substructures on the top right panel (the Drizzled image for 50 exposures). All the 15 substructures on the central galaxy can be found on the last panel, which is coadded by this work.}\label{CSST-MCI}
\end{figure}

%-------------------------------------------------------------------------%
\section{Simulation Tests based on the {\it CSST-MCI} mock pipeline}\label{sec_ST}
%-------------------------------------------------------------------------%

By employing the {\it CSST-MCI} mock pipeline, we take one HST Wide Field Camera 3 (WFC3) stamp from a real galaxy database of $Galsim$(\citealt{Rowe+2015}) as our mock input (or original image, ground truth), to generate mock exposures with random shifts and rotations for $g$-band. In the {\it CSST-MCI} mock, four blurring effects are taken into account: PSF convolution, cosmic ray, down-sampling, and Gaussian/Poisson noise. Following \cite{Wang+2022}, a non-strict positivity constraint is adopted:
\begin{equation}\label{cstr}
\sP_{\mathbb{R}^+}\{ \Tt^{(i+1)} \}=
\begin{cases}
\Tt^{(i+1)}&\Tt^{(i+1)} \ge 0\\
\Tt^{(i)}&\Tt^{(i+1)} < 0
\end{cases}
\end{equation}
where $\sP_{\mathbb{R}^+}\{ \Tt^{(i+1)} \}$ is a component-wise projection of $\Tt^{(i+1)}$ onto the set $\mathbb{R}^+$ (not strictly). For the convenience of comparison, we assumed that a PSF is uniform in a local region. The region size is determined by the spatial growth rate of the PSF. It is a local uniformity hypothesis for the PSFs.

\subsection{Mock-I: Faint sources detection}
This case centres on the application of mock and multi-exposure coaddition to {\it CSST-MCI}, i.e. Figure \ref{CSST-MCI}. The image in the lower left panel serves as the ground-truth input, extracted from an HST observation and showcasing no less than 15 substructures\footnote{The substructures of galaxies are detected by the software SourceXtractor++ (https://sourcextractorplusplus.readthedocs.io/en/latest/Introduction.html). It may be point-like sources or spiral arms in the figure.} on the central galaxy. The upper left panel features one of the $g$-band exposures ($512\times 512$ pixels, with a down-sampling factor $\beta =2$, convolved PSF size $128\times 128$ pixels), which displays two easily identifiable substructures. The $Drizzled$ image for 50 exposures is displayed in the upper right panel, revealing roughly seven substructures. Lastly, the final panel showcases the coadded image generated by this work, where all 15 substructures on the central galaxy are clearly visible. There are 7 stars in the field of view, corresponding to 7 pixels with high flux in the ground-truth input. The stars are visible in the rest panels but with different full half-maximum widths (FHWM).

\subsection{Mock-II: Strong lensing mock and images coaddition}

\begin{figure}
\centering
\includegraphics[width=6.1in]{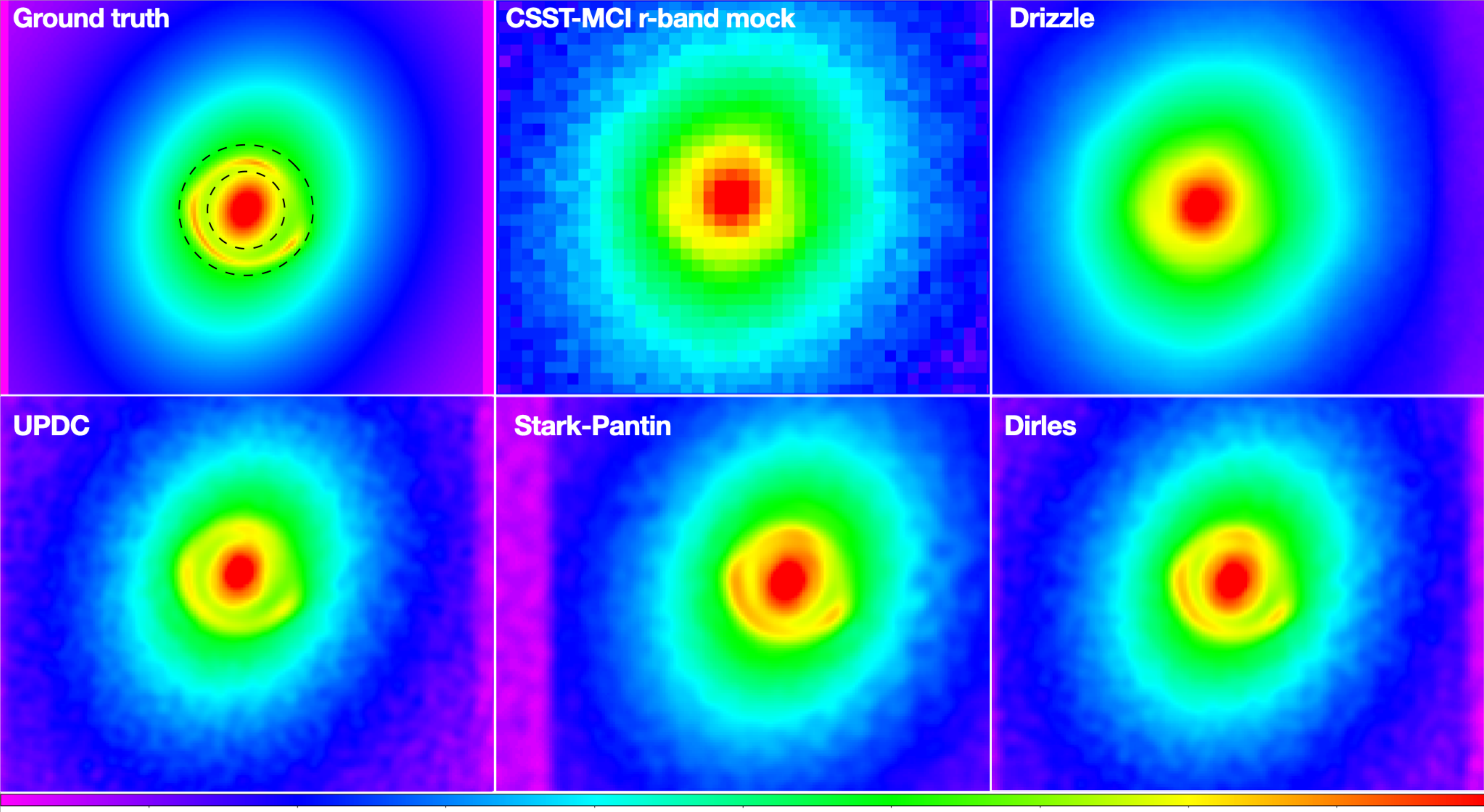} 
\caption{A strong gravitational lensing mock. The bottom left panel shows the ground-truth image as the mock input, which is generated by a lensing model. One of the {\it CSST-MCI} mocked exposures in $r$-band is illustrated on the top middle panel (with a down-sampling factor $\beta =4$). Four kinds of coaddition methods recovered images from exposures are shown on the rest panels: {\it Drizzle} (top right), {\it UPDC} (bottom left), {\it Starck-Pantin} (bottom middle)  and {\it Dirles} (bottom right).}\label{stronglense}
\end{figure}

Figure \ref{stronglense} shows a strong gravitational lensing mock and recoveries from four kinds of coaddition methods. The panel in the bottom left displays the ground-truth image generated by a lensing model and used as the mock input. Notably, the strong lensed giant arc within the dashed annulus is known as the {\it Einstein ring}. Moving to the top middle panel, we see an example of one of the {\it CSST-MCI} mocked exposures in the $r$-band with a down-sampling factor $\beta =4$. Finally, the remaining panels showcase four distinct coaddition methods that were used to recover images from the exposures ($128\times 128$ pixels, convolved PSF size $128\times 128$ pixels). These methods are referred to as {\it Drizzle} (top right), {\it UPDC} (bottom left), {\it Starck-Pantin} (bottom middle), and {\it Dirles} (bottom right). Visually, {\it UPDC} and {\it Dirles} manifest better results than the {\it Drizzle} or the {\it Starck-Pantin} in the {\it Einstein ring} recovery from 100 mocked exposures.

\begin{figure}
\centering
\includegraphics[width=4.1in]{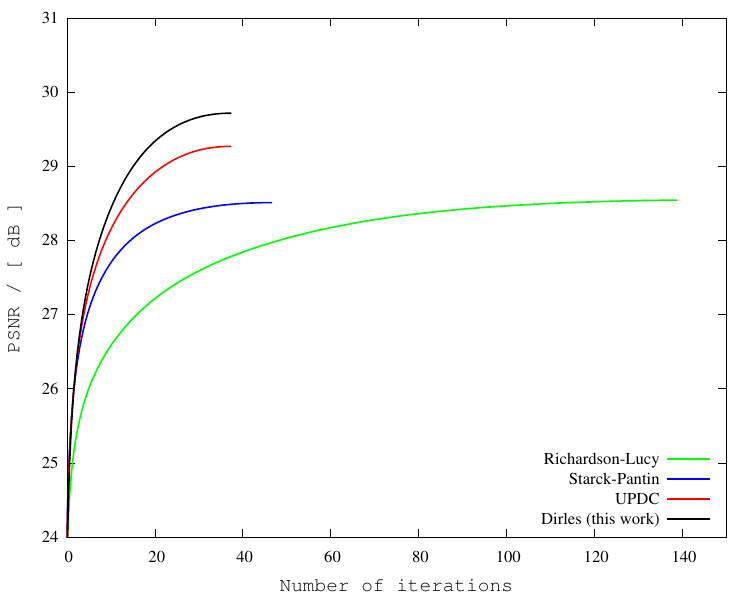} 
\caption{A comparison of the PSNR from different coaddition methods: {\it Drizzle} ($PSNR=22.16$), {\it UPDC} (in Red), {\it Richardson-Lucy} (in Green), {\it Starck-Pantin} (in Blue)  and {\it Dirles} (in Black). The curves end at the optimal iteration with the highest PSNR.}\label{figPNSR}
\end{figure}

Quantitatively, we measure the PSNR within the region of the dashed annulus in Figure \ref{stronglense} for each iteration of different coaddition methods. The result is shown in Figure \ref{figPNSR}. The methods tested are {\it Drizzle} (with a PSNR of 22.16), {\it UPDC} (in Red), {\it Richardson-Lucy} (in Green), {\it Starck-Pantin} (in Blue)  and {\it Dirles} (in Black). The curves end at the optimal iteration with the highest PSNR. The {\it Dirles} achieves the highest PSNR among all the recoveries with less than 40 iterations. {\it Richardson-Lucy} method consumes the most iterations but the least computation(see Table \ref{computationtime}). Note that the same PSFs are used in the PSF deconvolutions as those for the mock input.

\subsection{Mock-III: VST mock and image restoration}
To test the recovery methods, we have mocked the 100 frames $r$-band $VOICE-CDFS-1$ multi-exposures data from the OmegaCAM of the Very Large Telescope Survey Telescope in the European Southern Observatory. To reduce computation time, we extracted a square region that is centred at [$R.A.=53.1511^\circ, DEC.=-27.7175^\circ$] and has a size of $101\times101$ observation pixels. There is a high-resolution image (goodss\_3dhst\_F606W\_sci.fits) from the HST in this region. The high-resolution HST image has been binned $1.77858\times 1.77858$ times along the two dimensions.

\begin{figure}
\centering
\includegraphics[width=6.1in]{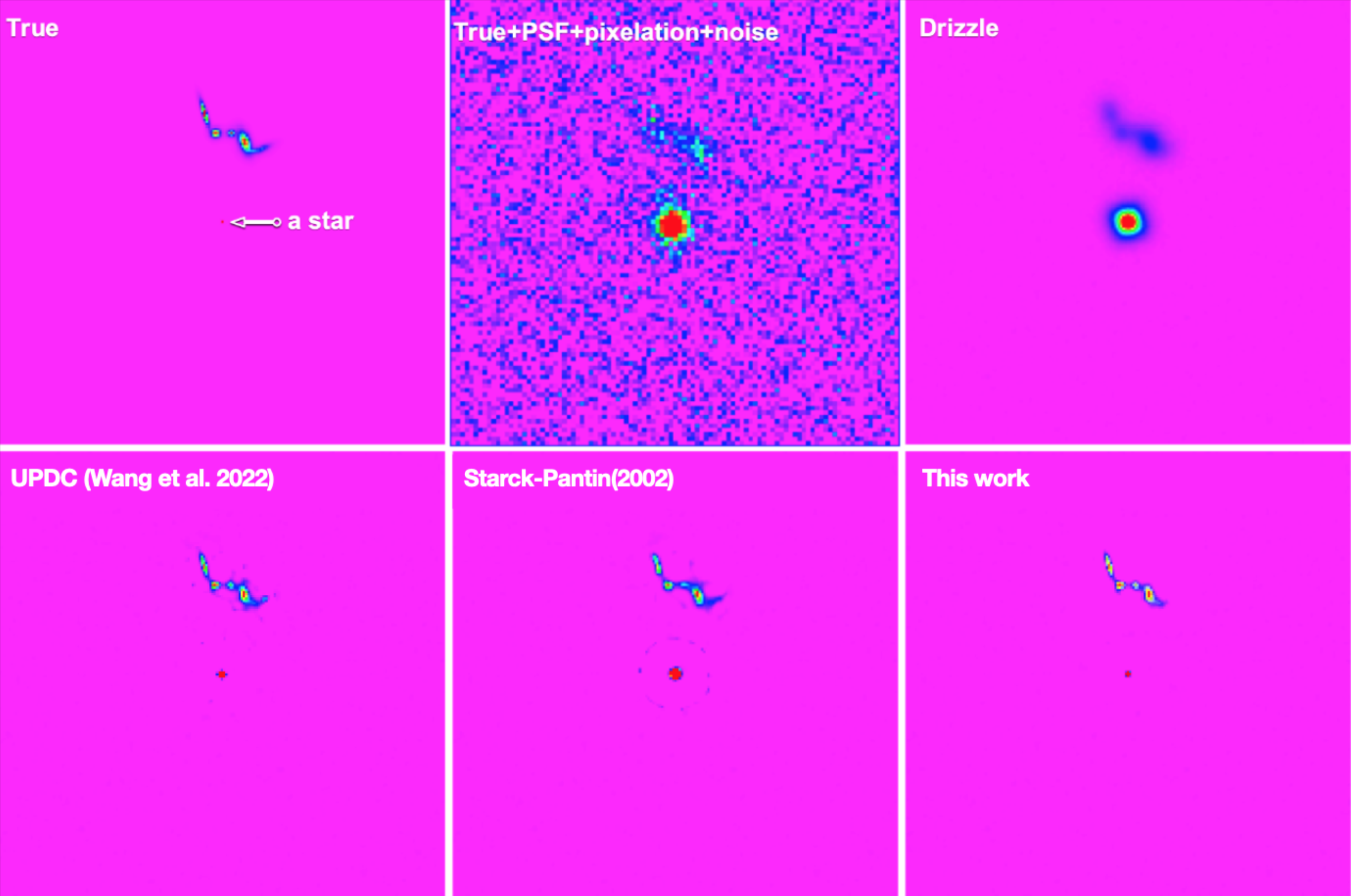} 
\caption{Recoveries from the VST mock. The top left panel displays the ground truth, which is a model fitting to the binned HST observation. One of the VST mock exposures is shown on the top middle panel. The top right panel displays the image recovered by using {\it Drizzle} method. The three panels at the bottom, from left to right, represent the recoveries by {\it UPDC}, {\it Starck-Pantin} and {\it Dirles} methods respectively.}
\label{VSTmock}
\end{figure}

In our simulation process, we use a group of Moffat profiles to replicate extended sources such as galaxies. To mimic the central point source (the star) on the target grid, we adopt a single pixel. By adjusting the Moffat profiles' parameters and the central point's brightness, and convolving a PSF from the binned HST image on the same sky field, we can fit a target image to the binned HST image. The target image is shown on the top left panel of Fig.~\ref{VSTmock}, and we consider it to be true.

To produce the mock samples, we first combine the fine PSFs and true PSFs measured in the 100 VOICE frames. The fine PSFs, which have a dimension of $59\times 59$ pixels, are generated by a Principal Component Analysis (PCA) based method developed by our team's cooperators \cite{Nie+2021a,Nie+2021b}. Next, we shift the image using the same position shifts as the VOICE frames and down-sample the true image to the same coarse grids as in the VOICE frames. Finally, we add the Poisson and Gaussian noise from each VOICE frame to the corresponding mock sample, as shown in the middle panel of the top array of Fig.~\ref{VSTmock}.

To make the results more realistic, we create sets of realizations by modifying the random seed in the Poisson and Gaussian processes. For each set of realizations, we reconstruct the mock samples using the methods mentioned in the text, namely {\it Drizzle}, {\it UPDC}, {\it Starck-Pantin}, and {\it Dirles}, up to the optimal number of iterations\footnote{The iteration has the highest PSNR.} in the $2\times2$ finer target grid.

According to the results depicted in Figure~\ref{VSTmock}, the {\it Dirles} approach appears to yield superior reconstructions in comparison to earlier methods. This can be attributed to its ability to effectively recover object shapes, minimize the ringing effect, and suppress background noise. It should be noted, however, that {\it Dirles} may disregard certain low SNR regions, such as the tails of galaxies. To circumvent this potential concern, decreasing the number of iterations and gradually balancing noise reduction with signal enhancement is advisable.

\begin{figure}
\centering
\includegraphics[width=6in]{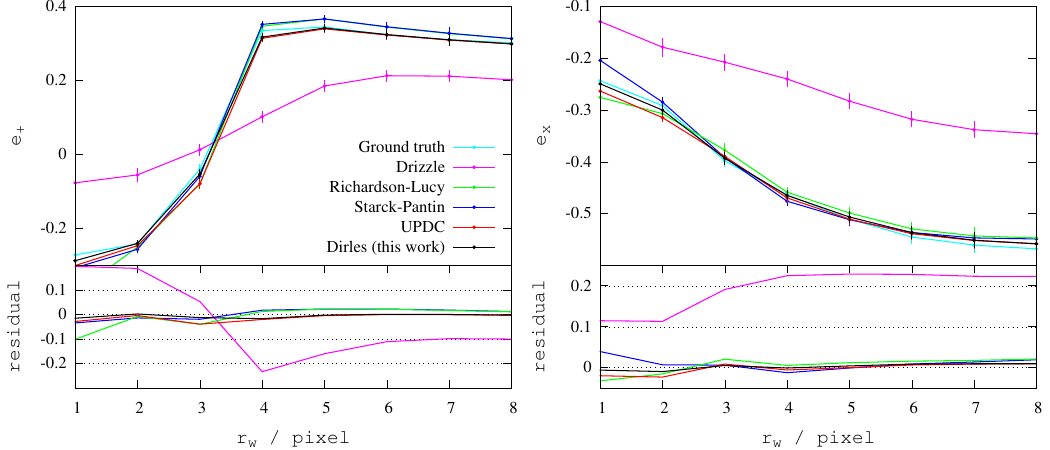} 
\caption{Shape parameters comparison between the true and the recoveries: left for the $e_+$ and right for the $e_\times$.}
\label{figshape}
\end{figure}

Weak gravitational lensing is primarily concerned with measuring the shear. This involves measuring the deformation of the background galaxies compared to the randomly aligned ones. By doing so, it is possible to constrain the properties of the foreground lensing objects. Therefore, the ellipticity of galaxies is the signal that researchers are interested in extracting. Following \cite{Hirata+2003}, the ellipticity of an object is defined as 
 \bea
e_+   &&= (M_{xx}-M_{yy})/(M_{xx}+M_{yy}) \nonumber\\
e_\times  &&= 2M_{xy}/(M_{xx}+M_{yy}) 
\label{eqedef}
\eea
where $M_{ij}$ represents the moments. the spin-2 tensor $\textbf{e} = (e_+,e_\times)$ is the so-called ellipticity tensor. To address the problem of divergence, we apply a circular Gaussian weighting function to the largest galaxy, with a weight radius of $r_w$. The variance of shape parameters is then plotted in Fig.~\ref{figshape} against the weight radius $r_w$. By analyzing the visual illustration in Fig.~\ref{VSTmock} and the quantitative results in Fig.~\ref{figshape}, we can conclude that the {\it Dirles} and {\it UPDC} methods outperform other methods in reconstructing shape parameters for extended sources. It is worth noting that the {\it Drizzle} method produced the largest deviation from the ground truth.

\begin{figure}
\centering
\includegraphics[width=6in]{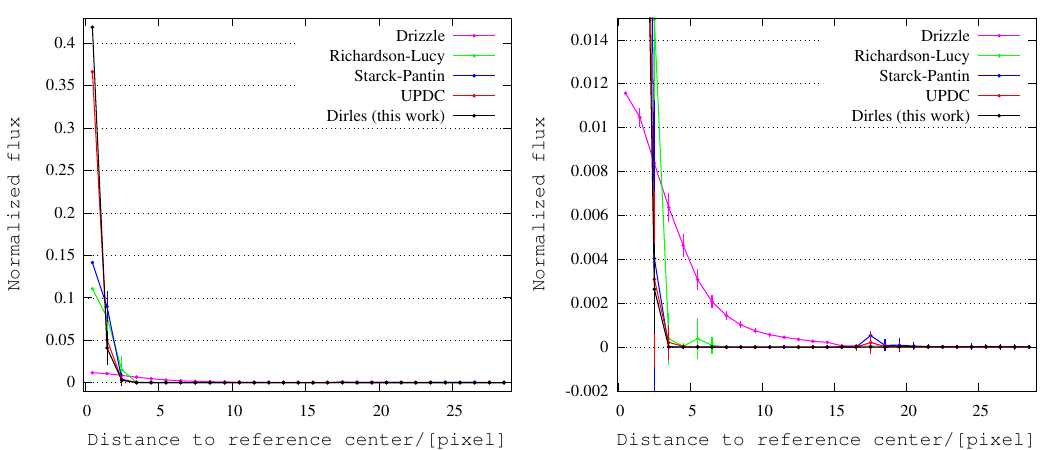} 
\caption{The left panel shows the normalized profiles of the central star drawn in different colours to differentiate between them: {\it Drizzle} (in Magenta), {\it UPDC} (in Red), {\it Richardson-Lucy} (in Green), {\it Starck-Pantin} (in Blue)  and {\it Dirles} (in Black). The right panel is a zoomed-in version that focuses on the area around the zero flux. This view highlights the ringing effect and other details around the point source.}
\label{PSFX}
\end{figure}

\begin{figure}
\centering
\includegraphics[width=5.1in]{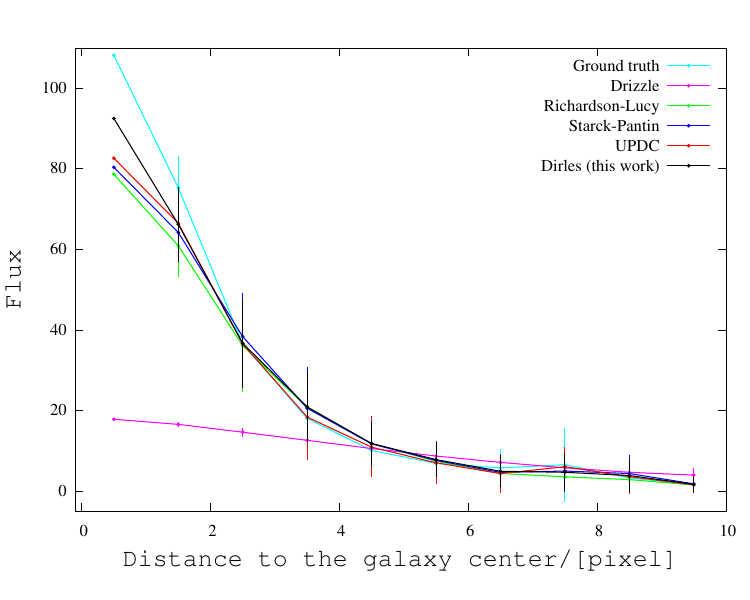} 
\caption{ Profiles of the largest galaxy from ground truth (in Cyan) and five recoveries: {\it Drizzle} (in Magenta), {\it UPDC} (in Red), {\it Richardson-Lucy} (in Green), {\it Starck-Pantin} (in Blue)  and {\it Dirles} (in Black) .}
\label{GalaxyProfile}
\end{figure}

In the field of photometry, the reconstructed source profile plays a pivotal role in estimating the performance of recovery algorithms. To analyze the profiles for both the central star and the largest nearby galaxy, we have presented graphical representations in Fig.~\ref{PSFX} and Fig.~\ref{GalaxyProfile}, respectively. Our analysis based on Fig.~\ref{VSTmock}, \ref{PSFX}, and Fig.~\ref{GalaxyProfile} demonstrates that {\it Dirles} outperforms other methods by effectively recovering high peak intensities in point sources, providing superior profiles for extended sources, and minimizing undesirable ringing artefacts. It is noteworthy that, for a given recovery method, the optimal number of iterations exhibits minimal variation across different realizations.

\begin{table}
\caption{Important Variables Used in the Text}\label{params}
    \centering
    \begin{tabular}[width=\columnwidth]{p{1.cm}||p{9cm}}
    \hline
    \bf{Variable} & \bf{Description} \\
    \hline
     $\Oo$ & intrinsic, continuous surface brightness of the sky\\
     $\Gg_k$ & the $k$-th exposure\\
     $\Tt$ & fully-sampled high resolution (or definition) image\\
     $\Pp_k$ & PSF measured from the $k$-th frame\\
    $\Ww$ &weight for flux allocation\\
    	$\mathbb{G}$ &sample for pixels on the same {\it WCS} position\\
	$\beta$ & down-sampling factor\\
	$\theta$ & noise at the pixel\\
     \hline
     $\bN\{\cdot\}$ & noise operator\\
     $\bB\{\cdot\}$ & blurring operator\\
    $\sP_{\mathbb{R}^+}\{\cdot\}$ & component-wise projection operator for constraints set\\
    \hline
    \end{tabular}
\end{table}

 \begin{table}
    \caption{Computational resources consumed by various coaddition approaches during each iteration}\label{computationtime}
        \centering
        \begin{tabular}{l||rrrrr|r}
        {\bf{Simu No.}} & {\it Drizzle} & {\it Richardson-Lucy}    & {\it Starck-Pantin}  & {\it UPDC}  & {\it Dirles}  \\
        \hline
        Mock-I ($512\times512\times50,\beta=2$) & 3.7s & 220.4ms & 7.9s  & 7.8s   & 272.2s \\
        Mock-II ($128\times128\times100,\beta=4$) & 1.9s & 90.1ms & 3.8s & 3.9s  & 144.3s \\
        Mock-III ($101\times101\times100,\beta=2$) & 0.3s & 6.8ms & 0.6s & 0.6s  & 5.4s
        \end{tabular}
    \end{table}
    
We have conducted a comprehensive study of the computation time required by different coaddition methods during each iteration. The results are presented in Table ~\ref{computationtime}. All calculations were performed on a two-socket AMD server equipped with 2 CPUs of EPYC 7763@2.45-3.5GHz and 1TB DDR4 memory. 

The {\it Richardson-Lucy} method has the lowest computation cost as it mainly executes the FFT in the calculation, which increases with an $O(N{\rm log}N)$ trend. On the other hand, {\it Drizzle} consumes the most time on the exposure up-sampling process. Both {\it Starck-Pantin} and {\it UPDC} involve similar operations such as PSF convolving (via FFT), down-sampling, comparing, and up-sampling. Hence, they have almost the same computation cost, which is mainly contributed by the sampling. 

{\it Dirles} has the highest computation cost compared to the other methods. Instead of the FFT-based convolution, its PSF convolution is performed in real space, which significantly increases computational time. Apart from the size of input exposures, the computation also depends on the dimension of the PSF provided in advance. However, the {\it Dirles} method stands out from other approaches as it does not assume local uniformity for the PSF. Consequently, it is a better fit for scenarios where the PSF undergoes significant changes within an exposure space. With {\it Dirles}, we can convolve a PSF map that is different everywhere in the field of view. Liu Dezi et al. (2023, in preparation) are constructing such a PSF map by using a set of basic functions for fitting.

%-------------------------------------------------------------------------%
\section{Conclusions}
\label{sec_DC}
%-------------------------------------------------------------------------%

In this article, we propose a novel AAPD method, named {\it Dirles}, which is based on the least squares fitting technique. Following a series of rigorous simulation tests conducted on these algorithms, our findings can be summarized as follows.

\begin{enumerate}
\item{In detecting faint sources, {\it Dirles} recovered all 15 substructures while {\it Drizzle} missed 8.}
 
\item{ The {\it Dirles} approach outperforms previous methods by achieving the highest PSNR, recovering object shapes, minimizing the ringing effect and suppressing noise. However, it may ignore low SNR regions. To address this, gradually reduce iterations and balance noise reduction with signal enhancement.}
  
\item{ The {\it Dirles} method outperforms other methods in reconstructing shape parameters for extended sources. In radiometry, the {\it Dirles} method performs the best in terms of recovering the highest peak in point sources, providing the best profile for extended sources, and reducing the ringing effect.}

\item{Compared to the other methods, {\it Dirles} has the highest computation cost. However, it stands out from other approaches as it does not assume local uniformity for the PSF, making it a better fit for scenarios where the PSF undergoes significant changes within an exposure space.}

\end{enumerate}

After extensive simulations, it can be concluded that {\it Dirles} outperforms previous works in restoring point/extended sources, maintaining extended source shapes, and reducing ringing despite high computation consumption.

In the future, numerous upcoming telescopes will commence astronomical observations, such as NASA's Wide Field Infrared Survey Telescope (WFIRST), the European Space Agency's Euclid mission, the National Science Foundation-funded Large Synoptic Survey Telescope (LSST), and China's Space Station Optical Telescope (CSST). These advanced instruments will generate a vast volume of imaging data. Effectively processing these images while maintaining high fidelity is an imminent necessity. As the implementation of {\it Dirles} algorithm uses a divide-and-conquer approach, it may be possible to accelerate it to a considerable extent using GPU-based parallel computing. We are confident that there is a significant potential for improving the speed and effectiveness of the {\it Dirles} method even further.
 
\section*{Acknowledgements}
We are grateful to the anonymous reviewers for their helpful comments, which greatly improved the presentation of this work. This work was supported by the GHfund A(202302017475). This work is also supported by the Foundation for Distinguished Young Scholars of Jiangsu Province (No. BK20140050), the National Natural Science Foundation of China (Nos. 11973070, 11333008, 11273061, 11825303, and 11673065), the China Manned Space Project with No. CMS-CSST-2021-A01, CMS-CSST-2021-A03, CMS-CSST-2021-B01 and the Joint Funds of the National Natural Science Foundation of China (No. U1931210). HYS acknowledges the support from Key Research Program of Frontier Sciences, CAS, Grant No. ZDBS-LY-7013 and Program of Shanghai Academic/Technology Research Leader. We acknowledge the support from the science research grants from the China Manned Space Project with CMS-CSST-2021-A04, CMS-CSST-2021-A07.

 In this study, a cluster is used with the SIMT accelerator made in China. The cluster includes many nodes each containing 2 CPUs and 4 accelerators. The accelerator adopts a GPU-like architecture consisting of a 16GB HBM2 device memory and many compute units. Accelerators connected to CPUs with PCI-E, the peak bandwidth of the data transcription between main memory and device memory is 16GB/s.

\bibliographystyle{raa}
\bibliography{bibtex}

\end{document}